\documentclass[aps,prc,twocolumn,superscriptaddress,nofootinbib,longbibliography,floatfix,10pt]{revtex4-1}

\usepackage{graphicx}
\usepackage{dcolumn}
\usepackage{bm}
\usepackage{morefloats}
\usepackage{multirow}
\usepackage{amssymb}
\usepackage{amsmath}
\usepackage{xcolor} 
\usepackage{longtable}
\usepackage{fix-cm}
\usepackage{mathptmx} % rm & math 
\usepackage[T1]{fontenc}
\usepackage[colorlinks,allcolors=blue]{hyperref}
\makeatletter
\def\NAT@def@citea{\def\@citea{\NAT@separator}}
\makeatother

\begin{document} 

\title{Merging of transport theory with TDHF: multinucleon transfer in U+U collisions}

\author{S. Ayik}\email{ayik@tntech.edu}
\affiliation{Physics Department, Tennessee Technological University, Cookeville, TN 38505, USA}
\author{B. Yilmaz}
\affiliation{Physics Department, Faculty of Science, Ankara University, 06100, Ankara, Turkey}
\author{O. Yilmaz}
\affiliation{Physics Department, Middle East Technical University, 06800, Ankara, Turkey}
\author{A. S. Umar}
\affiliation{Department of Physics and Astronomy, Vanderbilt University, Nashville, TN 37235, USA}
\date{\today}

\begin{abstract}  Multinucleon transfer mechanism in the collision of
${}^{238}\text{U}+{}^{238}\text{U}$ system is investigated at $E_\text{c.m.}
=833$~MeV in the framework of the quantal diffusion description based on the
stochastic mean-field approach (SMF).  Double cross-sections $\sigma(N,Z)$ as a
function of the neutron and proton numbers, the cross-sections $\sigma(Z)$ and
$\sigma(A)$ as a function of the atomic numbers and the mass numbers are
calculated for production of the primary fragments. The calculation indicates
the ${}^{238} \text{U}+{}^{238} \text{U}$ system may be located at an unstable
equilibrium state at the potential energy surface with a slightly negative
curvature along the beta stability line on the $(N,Z)-$plane. This behavior may
lead to rather large diffusion along the beta stability direction. 
\end{abstract}

%\pacs{24.10.Jv; 21.30.Fe; 21.65.-f; 26.60.-c}
 
\maketitle
 
\section{INTRODUCTION}
\label{sec1}
It has been recognized that multinucleon transfer in heavy-ion collisions involving
massive nuclei provide a suitable mechanism for synthesizing new neutron rich
heavy nuclei~\cite{adamian2003,zagrebaev2007,aritomo2009,adamian2010,barrett2015,devaraja2015,zhao2016,sekizawa2016,feng2017,sekizawa2017a,sekizawa2019,mun2019,saiko2019,jiang2020}. For this purpose, experimental investigations have been
carried out in heavy-ion collision with actinide targets near barrier energies
\cite{kozulin2012,kratz2013,watanabe2015,desai2019}. Collisions of massive
systems near barrier energies predominantly lead to dissipative deep-inelastic
reactions and quasi-fission reactions. In dissipative collisions the most part
of the bombarding energy is converted into the internal excitations, and the
multinucleon transfer occurs between the projectile and target nuclei. A number
of experimental and theoretical investigations have been made of the
multinucleon transfer mechanism in heavy-ion collisions near barrier energies.
The multi dimensional phenomenological Langevin type dynamical approaches have
been developed for describing dissipative collisions between massive nuclear
systems~\cite{zagrebaev2008c,zagrebaev2011,zagrebaev2012,karpov2017,saiko2019}.
These phenomenological models provide a qualitative and in some cases
semi-quantitative description of the transfer process. Since
many years, the time-dependent Hartree-Fock (TDHF) approach has been used for
describing the deep-inelastic collisions and the quasi-fission reactions
\cite{simenel2012,nakatsukasa2016,oberacker2014,oberacker2010,umar2015a,sekizawa2019,simenel2018}. The
TDHF provides a microscopic description in terms of Skyrme-type energy density
functionals. The mean-field theory provides good a description for the most
probable dynamical path of the collective motion at low energy heavy
ion-collisions including the one-body dissipation mechanism. However the
mean-field theory severely underestimates the fluctuations around the most
probable collective path. The particle number projection method of the TDHF
indeed shows the fragment mass and charge distributions are largely
underestimated for strongly damped collisions~\cite{simenel2010,sekizawa2016}.
The fragment mass and charge distributions observed in symmetric collisions
provide a good example for the shortcoming of the mean-field description. In the
TDHF calculations of the symmetric collisions, the identities of the projectile
and target are strictly preserved, i.e., the mass and charge numbers of the
final fragments are exactly same as of those at the initial fragments. The
experiments, on the other hand, exhibits broad mass and charge distributions of
final fragments around their initial values. The dominant aspect of the data is
a broad mass and charge distribution around the projectile and target resulting
from multinucleon diffusion mechanism. The description of such large
fluctuations requires an approach beyond the mean-field theory.  The
time-dependent RPA approach of Balian and Veneroni provides a possible approach
for calculating dispersion of fragment mass and charge distributions and
dispersion of other one-body observables
\cite{balian1984,balian1985,broomfield2008,williams2018,godbey2020}. However,
this approach has severe technical difficulties in applications to the
collisions of asymmetric systems. In this work, we employ the quantal diffusion
description based on the stochastic mean-field (SMF) approach to calculate
double cross-sections $\sigma (N,Z)$, the cross-section as function of mass
number $\sigma (A)$ and cross-section as a function of the atomic number $\sigma
(Z)$ of the primary fragments in the collisions of the symmetric ${}^{238}
\text{U}+{}^{238} \text{U}$ system at $E_\text{c.m.} =833$~MeV
\cite{ayik2008,lacroix2014}. In the quantal diffusion description, the transport
theoretical concepts are merged with the mean-field description of the TDHF. As
a result, it is possible to calculate the transport coefficients of macroscopic
variables in terms of the mean-field properties provided by the time-dependent
wave functions of the TDHF, which is consistent with the fluctuation-dissipation
theorem of the non-equilibrium statistical mechanics. In Sec.~\ref{sec2}, we present a
brief description of the quantal nucleon diffusion description of the
multinucleon exchange. In Sec.~\ref{sec3}, we present results of calculations of the
cross-sections for production of the primary fragments, and conclusions are
given in Sec.~\ref{sec4}. Some calculations details are provided in the Appendices.

\section{QUANTAL DIFFUSION OF MULTINUCLEON TRANSFERS}
\label{sec2}
\begin{figure}[!hpt]
\includegraphics*[width=8.6cm]{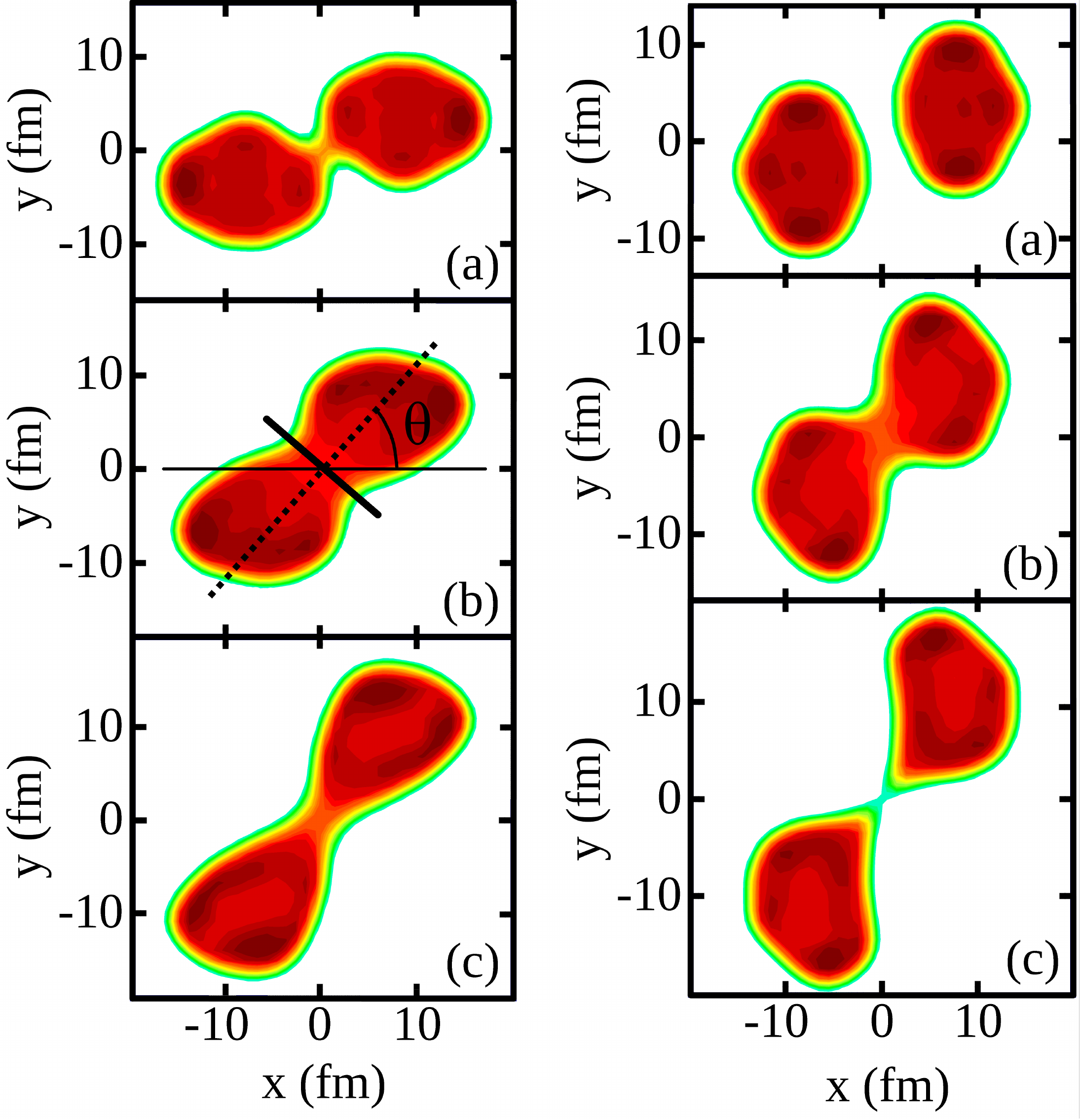}
\caption{The density profile and the collision geometry of the ${}^{238}\text{U}+{}^{238} \text{U}$ collisions at $E_\text{c.m.} =833$~MeV with the initial orbital angular momentum $\ell=300\hbar$ at times  $t=300$~fm/c , $t=500$~fm/c and $t=700$~fm/c at tip-tip geometry (left panel) and side-side geometry (right panel).}
\label{fig1}
\end{figure}

In the SMF approach, the dynamics of heavy-ion collisions is
described in terms of an ensemble of mean-field events. Each event is determined
by the self-consistent mean-field Hamiltonian of that event with the initial
conditions specified by the thermal and quantal fluctuations at the initial
state. We consider uranium-uranium collisions at bombarding energies near
Coulomb barrier. During the collision, the projectile and the target form a
di-nuclear complex and interact mainly by multinucleon exchanges. Because of
the di-nuclear structure, rather than generating an ensemble of stochastic
mean-field events, it is possible to describe the dynamics in terms of several
relevant macroscopic variables, such as neutron and proton numbers of the one
side of the complex and relative momentum of projectile-like and the target-like
fragments. It is possible to deduce the Langevin-type transport description for
the macroscopic variables~\cite{gardiner1991,weiss1999} and calculate transport
coefficients of the macroscopic variables in terms of the TDHF solutions. In
this manner, the SMF approach provides a ground for merging transport theory
with the mean-field description. For the detail description of the SMF
approach and the applications, we refer the reader the previous publications
\cite{ayik2017,ayik2018,yilmaz2018,ayik2019,sekizawa2020,yilmaz2020}. Here we
take the neutron $N_{1}^{\lambda }$ and the proton $Z_{1}^{\lambda}$ numbers of
the projectile-like fragments as the macroscopic variables. In each event
$\lambda$, the neutron and proton numbers are determined by integrating the
nucleon density over the projectile side of the window between the colliding
nuclei,
\begin{align} \label{eq1}
\left(\begin{array}{c} {N_{1}^{\lambda } (t)} \\ {Z_{1}^{\lambda } (t)} \end{array}\right)=\int d^{3} r\Theta \left[x'(t)\right]\left(\begin{array}{c} {\rho _{n}^{\lambda } (\vec{r},t)} \\ {\rho _{p}^{\lambda } (\vec{r},t)} \end{array}\right),
\end{align}
where $x'(t)=[y-y_{0} (t)]\sin\theta +[x-x_{0} (t)]\cos\theta$. The
$(x,y)$-plane represents the reaction plane with $x$-axis being the beam
direction in the center of mass frame (COM) of the colliding ions. The window plane
is perpendicular to the symmetry axis and its orientation is specified by the
condition $x'(t)=0$. In this expression, $x_{0}(t)$ and $y_{0}(t)$ denote the
coordinates of the window center relative to the origin of the COM frame,
$\theta(t)$ is the smaller angle between the orientation of the symmetry axis
and the beam direction.  We neglect fluctuations in the orientation of the
window and determine the mean evolution of the window dynamics by diagonalizing
the mass quadrupole moment of the system for each impact parameter $b$ or the
initial orbital angular momentum $\ell$, as described in Appendix~A, of Ref.
\cite{ayik2018}. In terms of the TDHF description, it is possible to determine
time evolution of the rotation angle $\theta (t)$ of the symmetry axis. The
coordinates $x_{0} (t)$ and $y_{0} (t)$ of the center point of the window are
located at the center of the minimum density slice on the neck between the
colliding ions. Since uranium is a deformed nucleus, the outcome of the
collisions depends on the relative orientation of the projectile and target. In
the present work, we consider two specific collision geometry: (i) the side-side
collisions in which deformation axes of the both the projectile and the target
are perpendicular to the beam direction and (ii) the tip-tip is collisions in
which deformation axes of the both the projectile and the target are parallel to
the beam direction. As an example, Fig.~\ref{fig1} shows the density profile in
the tip-tip geometry (left panel) and in the side-side geometry (right panel) of
the ${}^{238}\text{U}+{}^{238} \text{U}$ system at $E_\text{c.m.} =833$~MeV with
the initial orbital angular momentum $\ell=300\hbar$ at times $t=300$~fm/c ,
$t=500$~fm/c and $t=700$~fm/c. The window plane and symmetry axis of the
di-nuclear complex are indicated by thick and dashed lines in frame (b) of the left panel.
In the calculation of this figure and
in the calculations presented in the rest of the article, we employ the TDHF code developed by
Umar \textit{et al.}~\cite{umar1991a,umar2006c} using the SLy4d Skyrme functional~\cite{kim1997}.
In the following, all quantities are calculated for a given initial
orbital angular momentum $\ell$, but for the purpose of clarity of expressions,
we do not attach the angular momentum label to the quantities. The quantities in
Eq.~\eqref{eq1}
\begin{align} \label{eq2} 
 \rho _{\alpha }^{\lambda } (\vec{r},t)=\sum _{ij\in \alpha }\Phi _{j}^{*\alpha }  (\vec{r},t;\lambda )\rho _{ji}^{\lambda } \Phi _{i}^{\alpha } (\vec{r},t;\lambda )\;,
\end{align}
are the neutron and proton densities in the event $\lambda$ of the ensemble.
Here, and in the rest of the article, we use the notation $\alpha =n,p$ for the
neutron and proton labels.  According to the main postulate of the SMF approach,
the elements of the initial density matrix are specified by uncorrelated
Gaussian distributions with the mean values $\bar{\rho }_{ji}^{\lambda} =\delta
_{ji} n_{j}$ and the second moments determined by,
\begin{align} \label{eq3}
 \overline{\delta \rho _{ji}^{\lambda } \delta \rho _{i'j'}^{\lambda } }=\frac{1}{2} \delta _{ii'} \delta _{jj'} \left[n_{i} (1-n_{j} )+n_{j} (1-n_{i} )\right],
\end{align}
where $n_{j}$ are the average occupation numbers of the single-particle wave
functions at the initial state. At zero initial temperature, these occupation
numbers are zero or one, and at finite initial temperatures the occupation numbers
are given by the Fermi-Dirac functions. Here and below, the bar over the
quantity indicates the average over the generated ensemble. 

Below, we briefly discuss the derivation of the Langevin equations for the neutron and
proton numbers of the projectile-like fragments, for further details we refer the reader to
Refs.~\cite{ayik2017,ayik2018,yilmaz2018,ayik2019}. The rate of changes the neutron and the
proton numbers of the projectile-like fragment are given by,
\begin{align} \label{eq4}
\frac{d}{dt} \left(\begin{array}{c} {N_{1}^{\lambda } (t)} \\ {Z_{1}^{\lambda } (t)} \end{array}\right)=-\int d^{3} r\Theta \left[x'(t)\right] \left(\begin{array}{c} {\vec{\nabla }\cdot \vec{j}_{n}^{\lambda } (t)} \\ {\vec{\nabla }\cdot \vec{j}_{p}^{\lambda } (t)} \end{array}\right).
\end{align}
In obtaining this expression we neglect a term arising from the rate of change
of the position and the rotation of the window plane and employ the continuity
equation, with the fluctuating neutron and proton current densities
\begin{align} \label{eq5}
\vec{j}_{\alpha }^{\lambda } (\vec{r},t)==\frac{\hbar }{m} \sum _{ij\in \alpha }\text{Im}\left(\Phi _{j}^{*\alpha } (\vec{r},t;\lambda )\vec{\nabla }\Phi _{i}^{\alpha } (\vec{r},t;\lambda )\rho _{ji}^{\lambda } \right).
\end{align}
By carrying out a partial integration, we obtain a set of coupled
Langevin equations for the macroscopic variables $N_{1}^{\lambda} (t)$ and
$Z_{1}^{\lambda}(t)$,
\begin{align} \label{eq6}
\frac{d}{dt} \left(\begin{array}{c} {N_{1}^{\lambda } (t)} \\ {Z_{1}^{\lambda } (t)} \end{array}\right)=\int d^{3} rg(x')\left(\begin{array}{c} {\hat{e}\cdot \vec{j}_{n}^{\lambda } (\vec{r},t)} \\ {\hat{e}\cdot \vec{j}_{p}^{\lambda } (\vec{r},t)} \end{array}\right)=\left(\begin{array}{c} {v_{n}^{\lambda } (t)} \\ {v_{p}^{\lambda } (t)} \end{array}\right), 
\end{align}
with $\hat{e}$ as the unit vector along the symmetry axis with components
$\hat{e}_{x} =\cos \theta$ and $\hat{e}_{y} =\sin\theta$. In the integrand, we
replace the delta function by a smoothing function $\delta (x')\to g(x')$ in
terms of a Gaussian $g(x)=\left(1/\kappa \sqrt{2\pi} \right)\exp\left(-x^{2}
/2\kappa ^{2} \right)$ with dispersion $\kappa$. The Gaussian behaves almost
like delta function for sufficiently small $\kappa$. In the numerical
calculations dispersion of the Gaussian is taken in the order of the lattice
side $\kappa =1.0$ ~fm. The right side of Eq.~\eqref{eq6} defines the
fluctuating drift coefficients $v_{\alpha }^{\lambda } (t)$ for the neutrons and
the protons. There are two different sources for fluctuations of the drift
coefficients: (i) Fluctuations due to different set of wave functions in each
event $\lambda$. This part of the fluctuations can be approximately described in
terms of the fluctuating macroscopic variables as $v_{\alpha }^{\lambda } (t)\to
v_{\alpha } \left(N_{1}^{\lambda } (t),Z_{1}^{\lambda } (t)\right)$, and (ii)
fluctuations introduced by the stochastic part $\delta \rho _{ji}^{\lambda }
=\rho _{ji}^{\lambda } -\delta _{ji} n_{j}$ of the density matrix at the initial
state. In this work, we consider small amplitude fluctuations, and linearize the
Langevin Eq.~\eqref{eq6} around the mean values of the macroscopic variables
$\delta N_{1}^{\lambda } =N_{1}^{\lambda } -\overline{N}_{1}$ and $\delta
Z_{1}^{\lambda } =Z_{1}^{\lambda } -\overline{Z}_{1} $. The mean values
$\overline{N}_{1} =\overline{N_{1}^{\lambda } }$ and $\overline{Z}_{1}
=\overline{Z_{1}^{\lambda } }$ are determined by the mean-field description of
the TDHF approach. Table~\ref{tab1} and table~\ref{tab2} show the results of the
TDHF calculations for the mean values for a set of observable quantities in the
collisions of ${}^{238} \text{U}+{}^{238} \text{U}$ system at
$E_\text{c.m.}=833$~MeV for the range initial orbital angular momentum
$\ell=(100-460)\hbar$. 
\begin{table}[h]
\caption{Result of TDHF calculations for tip-tip collisions of ${}^{238}
\text{U}+{}^{238} \text{U}$ system at $E_\text{c.m.} =833$~MeV for final values
of mass and charge of the projectile-like ($A_{1}^{f}$, $Z_{1}^{f}$) and
target-like fragments ($A_{2}^{f}$, $Z_{2}^{f}$) final orbital angular momentum
$\ell_{f}$, total kinetic energy (TKE), total excitation energy $E*$, center of
mass $\theta _\text{c.m.} $ and laboratory scattering angles
($\theta_{1}^{\text{lab}} $,$\theta_{2}^{\text{lab}}$) for a set initial orbital
angular momentum $\ell_{i}$.}
\label{tab1}
\begin{ruledtabular}
\begin{tabular}{c c c c c c c c c c c }
$\ell_i\,$($\hbar$) & A$_1^f$ & Z$_1^f$ & A$_2^f$ & Z$_2^f$ & $\ell_f\,$($\hbar$) & TKE
& E$^*$ & $\theta_{c.m.}$ & $\theta_1^{lab}$ & $\theta_2^{lab}$ \\
&&&&&&(MeV)&(MeV)&&& \\
\hline
\rule{0pt}{3ex}
100 & 238 & 92.0 & 238 & 92.0 & 73.4 & 527 & 306 & 158 & 48.3 & 9.55 \\
120 & 238 & 92.0 & 238 & 92.0 & 95.4 & 514 & 319 & 154 & 49.6 & 11.5 \\
140 & 238 & 92.0 & 238 & 92.0 & 114 & 505 & 328 & 149 & 50.4 & 13.6 \\
160 & 238 & 92.0 & 238 & 92.0 & 132 & 521 & 312 & 149 & 51.7 & 13.7 \\
180 & 238 & 92.0 & 238 & 92.0 & 153 & 510 & 323 & 138 & 51.3 & 18.5 \\
200 & 238 & 92.3 & 238 & 91.7 & 172 & 515 & 317 & 132 & 50.8 & 20.9 \\
220 & 238 & 92.0 & 238 & 92.0 & 177 & 525 & 318 & 129 & 50.4 & 22.4 \\
240 & 238 & 92.0 & 238 & 92.0 & 182 & 552 & 281 & 126 & 51.6 & 24.1 \\
260 & 238 & 91.6 & 238 & 92.4 & 185 & 577 & 256 & 123 & 52.2 & 25.6 \\
280 & 238 & 92.0 & 238 & 92.0 & 189 & 595 & 238 & 120 & 51.6 & 27.4 \\
300 & 238 & 92.0 & 238 & 92.0 & 201 & 616 & 217 & 116 & 51.2 & 29.3 \\
320 & 238 & 92.0 & 238 & 92.0 & 225 & 625 & 208 & 113 & 50.1 & 31.0 \\
340 & 238 & 92.0 & 238 & 92.0 & 245 & 645 & 188 & 109 & 49.5 & 32.8 \\
360 & 238 & 92.0 & 238 & 92.0 & 271 & 654 & 179 & 106 & 48.3 & 34.6 \\
380 & 238 & 92.0 & 238 & 92.0 & 333 & 714 & 119 & 101 & 47.8 & 37.7 \\
400 & 238 & 92.0 & 238 & 92.0 & 374 & 751 & 82.3 & 98.4 & 47.5 & 39.5 \\
420 & 238 & 92.0 & 238 & 92.0 & 429 & 797 & 35.9 & 96.2 & 47.4 & 41.4 \\
440 & 238 & 92.0 & 238 & 92.0 & 439 & 785 & 48.1 & 93.4 & 45.8 & 42.5 \\
460 & 238 & 92.0 & 238 & 92.0 & 491 & 819 & 14.1 & 91.4 & 45.5 & 44.1 
\end{tabular}
\end{ruledtabular}
\end{table}
\begin{table}[h]
\caption{Same as Table~\ref{tab1} for side-side collisions.}
\label{tab2}
\begin{ruledtabular}
\begin{tabular}{c c c c c c c c c c c }
$\ell_i\,$($\hbar$) & A$_1^f$ & Z$_1^f$ & A$_2^f$ & Z$_2^f$ & $\ell_f\,$($\hbar$) & TKE
& E$^*$ & $\theta_{c.m.}$ & $\theta_1^{lab}$ & $\theta_2^{lab}$ \\
&&&&&&(MeV)&(MeV)&&& \\
\hline
\rule{0pt}{3ex}
100 & 238 & 92.3 & 238 & 91.7 & 71.3 & 658 & 173 & 154 & 62.3 & 12.5 \\
120 & 238 & 92.0 & 238 & 92.0 & 86.5 & 660 & 173 & 149 & 62.7 & 14.4 \\
140 & 238 & 92.0 & 238 & 92.0 & 99.2 & 660 & 173 & 145 & 62.0 & 16.5 \\
160 & 238 & 92.0 & 238 & 92.0 & 116 & 668 & 165 & 140 & 61.3 & 19.0 \\
180 & 238 & 92.0 & 238 & 92.0 & 140 & 659 & 174 & 134 & 59.3 & 21.2 \\
200 & 238 & 92.1 & 238 & 91.9 & 164 & 666 & 167 & 129 & 57.8 & 23.9 \\
220 & 238 & 92.0 & 238 & 92.0 & 184 & 673 & 160 & 125 & 57.6 & 26.0 \\
240 & 238 & 92.0 & 238 & 92.0 & 196 & 673 & 160 & 122 & 55.5 & 27.4 \\
260 & 238 & 92.0 & 238 & 92.0 & 214 & 682 & 214 & 119 & 52.2 & 28.2 \\
280 & 238 & 92.0 & 238 & 92.0 & 234 & 692 & 141 & 115 & 53.4 & 30.7 \\
300 & 238 & 92.0 & 238 & 92.0 & 258 & 699 & 134 & 112 & 52.1 & 32.5 \\
320 & 238 & 92.0 & 238 & 92.0 & 282 & 705 & 128 & 108 & 50.8 & 34.2 \\
340 & 238 & 92.0 & 238 & 92.0 & 302 & 713 & 120 & 105 & 49.8 & 35.6 \\
360 & 238 & 92.0 & 238 & 92.0 & 318 & 723 & 110 & 103 & 49.1 & 36.8 \\
380 & 238 & 92.0 & 238 & 92.0 & 333 & 736 & 96.6 & 102 & 48.6 & 37.8 \\
400 & 238 & 92.0 & 238 & 92.0 & 331 & 751 & 81.7 & 99.7 & 48.1 & 38.9 \\
420 & 238 & 92.0 & 238 & 92.0 & 373 & 768 & 65.0 & 97.8 & 47.6 & 40.1 \\
440 & 238 & 92.0 & 238 & 92.0 & 393 & 785 & 47.6 & 96.2 & 47.2 & 41.2 \\
460 & 238 & 92.0 & 238 & 92.0 & 410 & 802 & 31.4 & 94.9 & 46.9 & 42.1 
\end{tabular}
\end{ruledtabular}
\end{table}

The fluctuations evolve according to the linearized coupled Langevin equations, 
\begin{align} \label{eq7}
\frac{d}{dt} \left(\begin{array}{c} {\delta Z_{1}^{\lambda } } \\ {\delta N_{1}^{\lambda } } \end{array}\right)=&\left(\begin{array}{c} {\frac{\partial v_{p} }{\partial Z_{1} } (Z_{1}^{\lambda } -\overline{Z}_{1} )+\frac{\partial v_{p} }{\partial N_{1} } (N_{1}^{\lambda } -\overline{N}_{1} )} \\ {\frac{\partial v_{n} }{\partial Z_{1} } (Z_{1}^{\lambda } -\overline{Z}_{1} )+\frac{\partial v_{n} }{\partial N_{1} } (N_{1}^{\lambda } -\overline{N}_{1} )} \end{array}\right)\nonumber\\
&+\left(\begin{array}{c} {\delta v_{p}^{\lambda } (t)} \\ {\delta v_{p}^{\lambda } (t)} \end{array}\right), 
\end{align}
where the derivatives of drift coefficients are evaluated at the mean values
$\overline{N}_{1}$ and $\overline{Z}_{1}$. The linear limit provides a good
approximation for small amplitude fluctuations and it becomes even better if the
driving potential energy has nearly harmonic behavior around the mean values.
The stochastic part $\delta v_{\alpha }^{\lambda } (t)$ of drift coefficients
given by, 
\begin{align} \label{eq8}
\delta v_{\alpha }^{\lambda } (t)=\frac{\hbar }{m} \sum _{ij\in \alpha }\int d^{3} rg(x')\text{Im}\left(\Phi _{j}^{*\alpha } (\vec{r},t)\stackrel{\frown}{e}\cdot \vec{\nabla }\Phi _{i}^{\alpha } (\vec{r},t)\delta \rho _{ji}^{\lambda } \right).
\end{align}

According to the basic postulate of the SMF approach the stochastic elements of
the initial density matrix $\delta \rho _{ji}^{\lambda }$ are specified in terms
of uncorrelated distributions, then it follows that the stochastic part of the
neutron and proton drift coefficients $\delta v_{\alpha }^{\lambda } (t)$ are
determined by uncorrelated Gaussian distributions with variances discussed in the
following section.

\section{MASS AND CHARGE DISTRIBUTIONS OF THE PRIMARY FRAGMENTS}
\label{sec3}

\subsection{Quantal diffusion coefficients of neutrons and protons}
\begin{figure}[!hpt]
\includegraphics*[width=8.6cm]{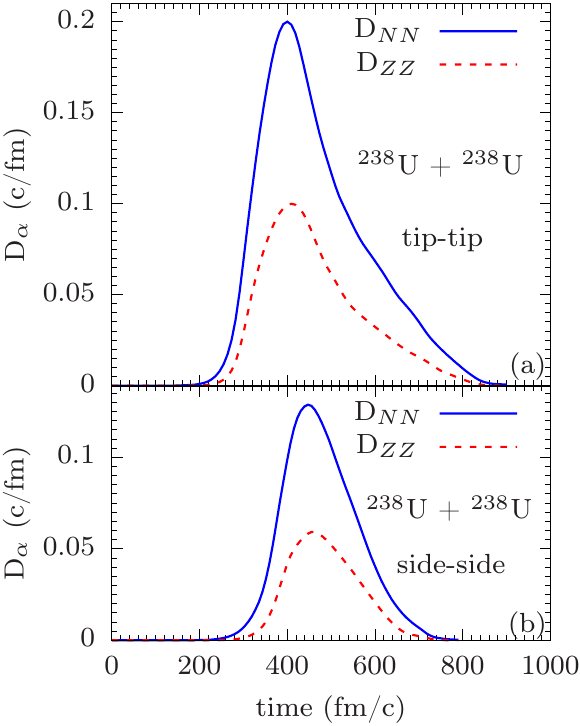}
\caption{Neutron and proton diffusion coefficients as a function of time in the
${}^{238} \text{U}+{}^{238} \text{U}$ collisions at $E_\text{c.m.} =833$ ~MeV
with the initial orbital angular momentum $\ell=300\hbar $ at tip-tip geometry
(a) and side-side geometry (b).}
\label{fig2}
\end{figure}

It is well known that Langevin equation for a macroscopic variable is equivalent
to the Fokker-Planck equation for the distribution function of the macroscopic
variable and the solution is given by a single Gaussian function
\cite{risken1996}.  When there are two coupled Langevin equations, as we have it
in Eq.~\eqref{eq7}, the solution of the Fokker-Planck equation for the
distribution function $P\left(N,Z\right)$ of fragments with neutron and proton
numbers $\left(N,Z\right)$ is specified by a correlated Gaussian function for
each value of the initial orbital angular momentum $\ell$,
\begin{align} \label{eq9}
P(N,Z)=\frac{1}{2\pi \sigma _{NN} \sigma _{ZZ} \sqrt{1-\rho ^{2} } } \exp \left[-C\left(N,Z\right)\right]. 
\end{align}
Here the exponent $C\left(N,Z\right)$ is given by
\begin{align} \label{eq10} 
&\overline{C}\left(N,Z\right)=\frac{1}{2\left(1-\rho^{2}\right)}\nonumber\\
&\times \left[\left(\frac{Z-\overline{Z}}{\sigma_{ZZ}} \right)^{2} -2\rho\left(\frac{Z-\overline{Z}}{\sigma_{ZZ}}\right)\left(\frac{N-\overline{N}}{\sigma_{NN}}\right)+\left(\frac{N-\overline{N}}{\sigma_{NN}}\right)^{2} \right],
\end{align}  
with the correlation coefficient $\rho =\sigma_{NZ}^{2} /\sigma _{ZZ} \sigma
_{NN}$. In this expression $\overline{N}$ and $\overline{Z}$ are the mean values
the neutron and the proton numbers of fragments for each angular momentum
determined by the TDHF calculations, and $\sigma _{NN}$, $\sigma _{ZZ} $ and
$\sigma _{NZ}$ denote the neutron, proton and mixed dispersions,
respectively. Multiplying both side in Eq.~\eqref{eq7} by $\delta
N_{1}^{\lambda}$ and $\delta Z_{1}^{\lambda}$ and carrying out ensemble
averaging, we obtain a couple set of equations for the neutron $\sigma _{NN}^{2}
=\overline{\delta N_{1}^{\lambda } \delta N_{1}^{\lambda}}$, the proton $\sigma
_{ZZ}^{2} =\overline{\delta Z_{1}^{\lambda } \delta Z_{1}^{\lambda}}$ and the
mixed variances $\sigma _{NZ}^{2} =\overline{\delta N_{1}^{\lambda } \delta
Z_{1}^{\lambda}}$, where bar indicates the ensemble averaging
\cite{schroder1981,merchant1982},
\begin{align} \label{eq11}
\frac{\partial }{\partial t} \sigma _{NN}^{2} =2\frac{\partial v_{n} }{\partial N_{1} } \sigma _{NN}^{2} +2\frac{\partial v_{n} }{\partial Z_{1} } \sigma _{NZ}^{2} +2D_{NN}, 
\end{align}
\begin{align} \label{eq12}
\frac{\partial }{\partial t} \sigma _{ZZ}^{2} =2\frac{\partial v_{p} }{\partial Z_{1} } \sigma _{ZZ}^{2} +2\frac{\partial v_{p} }{\partial N_{1} } \sigma _{NZ}^{2} +2D_{ZZ},
\end{align}
and
\begin{align} \label{eq13}
\frac{\partial }{\partial t} \sigma _{NZ}^{2} =\frac{\partial v_{p} }{\partial N_{1} } \sigma _{NN}^{2} +\frac{\partial v_{n} }{\partial Z_{1} } \sigma _{ZZ}^{2} +\sigma _{NZ}^{2} \left(\frac{\partial v_{p} }{\partial Z_{1} } +\frac{\partial v_{n} }{\partial N_{1} } \right). 
\end{align}
In these expressions $D_{NN}$ and $D_{ZZ}$ denote the neutron and proton quantal
diffusion coefficients which are discussed below. The expression of the
diffusion coefficients of for neutron and proton transfers are determined by the
auto-correlation functions of the stochastic part of the drift coefficients as
\begin{align} \label{eq14}
\int _{0}^{t}dt'\overline{\delta v_{\alpha }^{\lambda } (t)\delta v_{\alpha }^{\lambda } (t')} =D_{\alpha \alpha } (t).
\end{align} 
We can calculate the ensemble averaging by employing the basic postulate of the
SMF approach given by Eq.~\eqref{eq3}. We refer reader to Refs.
\cite{ayik2017,ayik2018} in which a detailed description of the autocorrelation
functions are presented. Here, for completeness of the presentation, we give the
results for the quantal expression of the proton and the neutron diffusion
coefficients,
\begin{align} \label{eq15}
D_{\alpha\alpha}(t)= & \int_{0}^{t}d\tau\int d^{3} r\,\tilde{g}(x')\left[G_{T}(\tau)J_{\bot,\alpha}^{T}(\vec{r},t-\tau/2)\right.\nonumber\\
 &\qquad\qquad\qquad\quad\left.+G_{P}(\tau)J_{\bot ,\alpha }^{P} (\vec{r},t-\tau /2)\right]\nonumber\\
&-\int_{0}^{t}d\tau\,\text{Re}\!\left(\sum_{h'\in P,h\in T}A_{h'h}^{\alpha}(t)A_{h'h}^{*\alpha}(t-\tau)\right.\nonumber\\
&\qquad\qquad\quad\left.+\sum_{h'\in T,h\in P}A_{h'h}^{\alpha}(t)A_{h'h}^{*\alpha}(t-\tau)\right).
\end{align} 
Here $J_{\bot ,\alpha }^{T} (\vec{r},t-\tau /2)$ represents the sum of the magnitude of current densities perpendicular to the window due to the hole wave functions originating from target,
\begin{align} \label{eq16}
J_{\bot ,\alpha }^{T} (\vec{r},t-\tau /2)=\frac{\hbar }{m} \sum _{h\in T}&\left|\frac{}{}\text{Im}\left[\Phi _{h}^{*\alpha } (\vec{r},t-\tau /2)\right.\right.\nonumber\\
&\left.\left.\times\left(\hat{e}\cdot \vec{\nabla }\Phi _{h}^{\alpha} (\vec{r},t-\tau /2)\right)\right]\right|.
\end{align}
and $J_{\bot ,\alpha }^{P} (\vec{r},t-\tau /2)$ is given by a similar expression
in terms of the hole wave functions originating from the projectile. We observe
that there is a close analogy between the quantal expression and the diffusion
coefficient in a random walk problem~\cite{gardiner1991,weiss1999}. The first
line in the quantal expression gives the sum of the nucleon currents across the
window from the target-like fragment to the projectile-like fragment and from
the projectile-like fragment to the target-like fragment, which is integrated
over the memory. This is analogous to the random walk problem, in which the
diffusion coefficient is given by the sum of the rate for the forward and
backward steps.  The second line in the quantal diffusion expression stands for
the Pauli blocking effects in nucleon transfer mechanism, which does not have a
classical counterpart. The quantities in the Pauli blocking factors are
determined by
\begin{align} \label{eq17}
A_{h'h}^{\alpha } (t)=\frac{\hbar }{2m} \int d^{3} r g(x')&\left(\Phi _{h'}^{*\alpha } (\vec{r},t)\hat{e}\cdot \vec{\nabla }\Phi _{h}^{\alpha } (\vec{r},t)\right.\nonumber\\
&\left.\quad-\Phi _{h}^{\alpha } (\vec{r},t)\hat{e}\cdot \vec{\nabla }\Phi _{h'}^{*\alpha } (\vec{r},t)\right).
\end{align} 
The memory kernels $G_{T} (\tau )$ in Eq.~\eqref{eq15} is given by 
\begin{align} \label{eq18}
G_{T} (\tau )=\frac{1}{\sqrt{4\pi } } \frac{1}{\tau _{T} } \exp [-(\tau /2\tau _{T} )^{2} ]
\end{align} 
with the memory time determined by the average flow velocity $u_{T}$ of the
target nucleons across the window according to $\tau _{T} =\kappa /|u_{T} (t)|$,
and $G_{P} (\tau )$ is given by a similar expression. In a previous work
\cite{ayik2018}, we estimated the memory time to be about $\tau _{T} =\tau _{P}
\approx 25$~fm/c, which is much shorter than the contact time of about $600$~fm/c.
As a result the memory effect is not important in diffusion coefficients.
We note that the quantal diffusion coefficients are entirely determined in terms
of the occupied single-particle wave functions of the TDHF solutions. According
to the non-equilibrium fluctuation-dissipation theorem, the fluctuation
properties of the relevant macroscopic variables must be related to the mean
properties. Consequently, the evaluation of the diffusion coefficients in terms
of the mean-field properties is consistent with the fluctuation-dissipation
theorem. Fig.~\ref{fig2} shows neutron and proton  diffusion coefficients for
the ${}^{238} \text{U}+{}^{238} \text{U}$ system at $E_\text{c.m.} =833$~MeV
with the initial orbital angular momentum $\ell=300\hbar$, for the tip-tip (a)
and the side-side (b) geometries  as function of time. 

\begin{figure}[!hpt]
\includegraphics*[width=8.6cm]{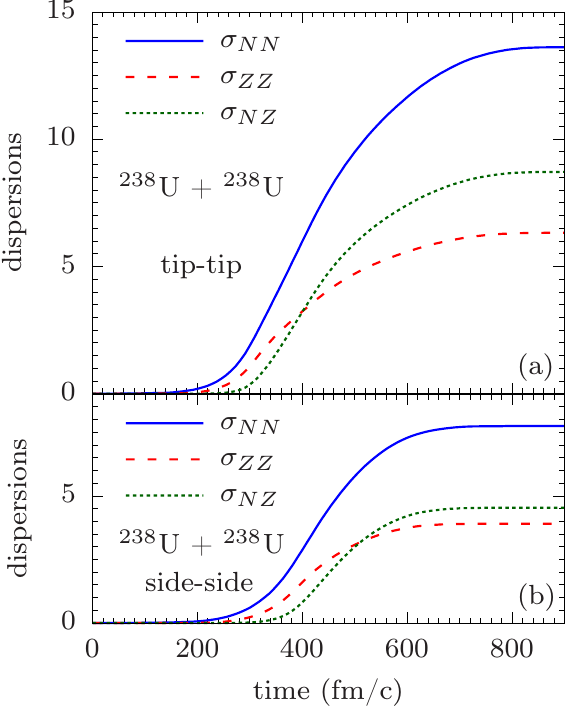}
\caption{Neutron, proton and mixed variance as a function of time in the
${}^{238} \text{U}+{}^{238} \text{U}$ collisions at $E_\text{c.m.} =833$~MeV
with the initial orbital angular momentum $\ell=300\hbar$ at tip-tip geometry
(a) and side-side geometry (b).}
\label{fig3}
\end{figure}

Dispersions are determined from the solutions of the coupled differential
equations (11-13) in which the diffusion coefficients provide source for
development of the fluctuations. In addition to the diffusion coefficients, we
also need to determine the derivatives of the drift coefficients with respect to
the macroscopic variables $\left(N_{1} ,Z_{1} \right)$. In order to determine
these derivatives, the Einstein's relations in the over-damped limit provide a
possible approach.  According to the Einstein relation, drift coefficients are
determine by the derivatives of the potential energy surface in the
$\left(N,Z\right)$-plane, 
\begin{align} \label{eq19}
 v_{n} (t)=&-\frac{D_{NN} }{T^{*} } \frac{\partial }{\partial N_{1} } U\left(N_{1} ,N_{1} \right)\nonumber\\
 v_{p} (t)=&-\frac{D_{ZZ} }{T^{*} } \frac{\partial }{\partial Z_{1} } U\left(N_{1} ,Z_{1} \right),
\end{align}
where $T^{*}$ indicates effective temperature of the system. Because of the
analytical structure, we can immediately take derivatives of the drift
coefficients. Since ${}^{238} \text{U}+{}^{238} \text{U}$ is a symmetric system,
the equilibrium state in the potential energy surface is located at the initial
position with $N_{1} \to N_{0} =146$ and $Z_{1} \to Z_{0} =92.$ When
fluctuations are not too far from the equilibrium point, we can parameterize the
potential energy around the equilibrium in terms of two parabolic forms as given
by Eq.~\eqref{a1} in the App.~\ref{appA}~\cite{merchant1982}. One of the parabolic
forms extend along the bottom of the beta stability line, which is referred to
as the iso-scalar path. The second parabolic form extends towards the
perpendicular direction to the iso-scalar path, which is referred to as the
iso-vector path. In order to specify the derivatives of the drift coefficients,
we need to determine the reduced curvature parameters $\alpha$ and $\beta$ of
these parabolic potential energy surfaces. Since the symmetric collisions do not
exhibit drift in neutron or proton numbers, it is not possible to specify the
reduced curvature parameters from the mean trajectory information of the
symmetric collisions. As discussed in App.~\ref{appA}, we can estimate the iso-vector
curvature $\alpha$ parameter from the central collision of the neighboring
${}_{88}^{236} \text{Ra}_{148} +{}_{96}^{240} \text{Cm}_{144}$ system at at
$E_\text{c.m.} =833$~MeV. As seen from the drift path of this system in
Fig.~\ref{figA1}, the system follows the iso-vector path closely and reaches the
charge equilibrium rather rapidly during a time interval of $\Delta t\approx
150$~fm/c.  The iso-vector drift path is suitable to estimate the average value
of the iso-vector curvature parameters and we find $\alpha \approx 0.13$. After
reaching the equilibrium in charge asymmetry rather rapidly, the system spends a
long time in the vicinity of ${}^{238} \text{U}+{}^{238} \text{U}$ by
following a curvy path due to complex quantal effect due to shell structure. 
Eventually, the system has a tendency to evolve toward asymmetry direction along
the iso-scalar path, i.e along the beta stability line. It appears that the
${}^{238} \text{U}+{}^{238} \text{U}$ system is located at an unstable state on
the beta stability line with a small and negative curvature parameter $\beta$ in
the iso-scalar direction. It is not possible to provide reasonable estimation
for this parameter from the drift path of the ${}_{88}^{236} \text{Ra}_{148}
+{}_{96}^{240} \text{Cm}_{144}$ system in Fig.~\ref{figA2} beyond the
equilibrium state at $(N_{1} =146,Z_{1} =92)$. With a negative curvature
parameter in the iso-scalar direction, the system may exhibit broad diffusion
along the beta stability line. In order obtain a reasonable value for $\beta$,
we employ the cross-section data for production of gold isotopes from a previous
investigation of the ${}^{238} \text{U}+{}^{238} \text{U}$ system at about the
same energy~\cite{kratz2013}. As discussed in Appendix~\ref{appB}, we determine a small negative value
of $\beta =-0.02$ for the reduced iso-scalar curvature parameter. Using these
values for the reduced curvature parameters, we can determine the derivative of
the drift coefficients as given in Eqs. \eqref{a5}-\eqref{a8} and calculate the
neutron, the proton and the mixed dispersions from the solution of the
differential Eqs. \eqref{eq11}-\eqref{eq13}. As an example Fig.~\ref{fig3} shows
the  neutron,  the proton and the mixed dispersions as a function of time in the
${}^{238} \text{U}+{}^{238} \text{U}$ collisions at $E_\text{c.m.} =833$~MeV
with the initial orbital angular momentum $\ell=300\hbar$ at tip-tip geometry
and side-side geometry. The asymptotic values of these dispersions for a range
of the initial orbital angular momentum $\ell=(100-460)\hbar$ in tip-tip and
side-side geometries are given in table \ref{tab3}. 

\begin{table}[!htb]
    \caption{Asymptotic values of the neutron, the proton and the mixed
dispersions in the ${}^{238} \text{U}+{}^{238} \text{U}$ collisions at
$E_\text{c.m.} =833$~MeV with the range of orbital angular momentum
$\ell=(100-460)\hbar$ at tip-tip geometry (left panel) and side-side geometry
(right panel).}
    \label{tab3}
    \vspace{0.2cm}
    \begin{minipage}{.5\linewidth}
       \centering
       \begin{tabular}{|m{0.65cm}|m{0.65cm}|m{0.65cm}|m{0.65cm}|m{0.65cm}|}
          \hline
          \multicolumn{5}{|c|}{$\alpha=0.13\;\; \beta=-0.02\;\,$ (tip-tip)} \\
          \hline
          $\ell_i(\hbar)$ & $\sigma_{NN}$ & $\sigma_{ZZ}$ & $\sigma_{NZ}$ & $\sigma_{AA}$ \\ 
          \hline
          100 & 21.7 & 9.91 & 14.3 & 31.3 \\
          \hline
          120 & 22.2 & 10.2 & 14.7 & 28.5 \\
          \hline
          140 & 23.0 & 10.6 & 15.3 & 29.6 \\
          \hline
          160 & 22.9 & 10.4 & 15.1 & 29.4 \\
          \hline
          180 & 23.2 & 10.5 & 15.3 & 29.7 \\
          \hline
          200 & 22.6 & 10.2 & 14.9 & 32.5 \\
          \hline
          220 & 21.4 & 9.68 & 14.0 & 30.8 \\
          \hline
          240 & 19.2 & 8.76 & 12.6 & 27.6 \\
          \hline
          260 & 14.3 & 7.97 & 11.3 & 25.0 \\
          \hline
          280 & 15.6 & 7.15 & 10.1 & 22.3 \\
          \hline
          300 & 13.6 & 6.32 & 8.71 & 19.4 \\
          \hline
          320 & 13.3 & 6.23 & 8.54 & 19.0 \\
          \hline
          340 & 11.6 & 5.53 & 7.37 & 16.6 \\
          \hline
          360 & 10.3 & 4.99 & 6.45 & 14.7 \\
          \hline
          380 & 6.97 & 3.53 & 3.89 & 9.55 \\
          \hline
          400 & 5.30 & 2.79 & 2.54 & 6.98 \\
          \hline
          420 & 3.46 & 1.72 & 1.05 & 4.15 \\
          \hline
          440 & 3.93 & 2.07 & 1.43 & 4.88 \\
          \hline
          460 & 2.49 & 1.14 & 0.50 & 2.83 \\
          \hline
       \end{tabular}
    \end{minipage}%
    \begin{minipage}{.5\linewidth}
       \centering
       \begin{tabular}{|m{0.7cm}|m{0.7cm}|m{0.7cm}|m{0.7cm}|m{0.7cm}|}
          \hline
          \multicolumn{5}{|c|}{$\alpha=0.13\;\; \beta=-0.02\,$ (side-side)} \\
          \hline
          $\ell_i(\hbar)$ & $\sigma_{NN}$ & $\sigma_{ZZ}$ & $\sigma_{NZ}$ & $\sigma_{AA}$ \\ 
          \hline
          100 & 12.5 & 5.86 & 7.95 & 16.0 \\
          \hline
          120 & 12.4 & 5.80 & 7.85 & 15.8 \\
          \hline
          140 & 12.1 & 5.71 & 7.69 & 15.5 \\
          \hline
          160 & 11.4 & 5.44 & 7.18 & 14.5 \\
          \hline
          180 & 11.4 & 5.43 & 7.19 & 14.5 \\
          \hline
          200 & 10.7 & 5.13 & 6.68 & 13.6 \\
          \hline
          220 & 10.0 & 4.86 & 6.22 & 12.7 \\
          \hline
          240 & 9.77 & 4.76 & 6.05 & 12.4 \\
          \hline
          260 & 9.04 & 4.45 & 5.51 & 11.5 \\
          \hline
          280 & 8.32 & 4.15 & 4.97 & 11.7 \\
          \hline
          300 & 7.75 & 3.91 & 4.53 & 10.8 \\
          \hline
          320 & 7.23 & 3.68 & 4.13 & 10.0 \\
          \hline
          340 & 6.74 & 3.45 & 3.73 & 9.23 \\
          \hline
          360 & 6.22 & 3.22 & 3.30 & 8.41 \\
          \hline
          380 & 5.63 & 2.95 & 2.81 & 7.50 \\
          \hline
          400 & 4.97 & 2.64 & 2.28 & 6.49 \\
          \hline
          420 & 4.26 & 2.26 & 1.70 & 5.39 \\
          \hline
          440 & 3.50 & 1.82 & 1.12 & 4.25 \\
          \hline
          460 & 2.82 & 1.37 & 0.68 & 3.28 \\
          \hline
       \end{tabular}
    \end{minipage}%
\end{table}

\subsection{Cross-section of production of primary fragments}
\label{xsec}
We calculate the cross-section for production of a primary fragment with neutron and proton numbers $(N,Z)$ using the standard expression,
\begin{align} \label{eq20}
\sigma (N,Z)=\frac{\pi \hbar ^{2} }{2\mu E_\text{c.m.} } \sum _{\ell_{\min } }^{\ell_{\max } }(2\ell+1)P_{\ell}  (N,Z).
\end{align} 
Here, $P_{\ell} (N,Z)=\left(P_{\ell}^{t-t} (N,Z)+P_{\ell}^{s-s} (N,Z)\right)/2$
denotes the mean value of the probability of producing a primary fragment with
neutron and proton numbers $(N,Z)$ in the tip-tip and the side-side collisions
with the initial angular momentum $\ell$. These probabilities are presented in
Eqs.~\eqref{eq9}-\eqref{eq10} with the asymptotic values of dispersions
given in Table~\ref{tab3} for tip-tip and side-side collisions. The mean values
are equal to their initial values $\overline{N}=146$, $\overline{Z}=92$. The
range of the summation over the initial angular momentum is taken as 
$\ell_{\min } =300$ and $\ell_{\max } =460$. This angular momentum range
corresponds the experimental set up in which the detector is placed at an angular
range $\theta =35^{\circ} \mp 5^{\circ}$ in the laboratory frame.
\begin{figure}[!hpt]
	\includegraphics*[scale=1.2]{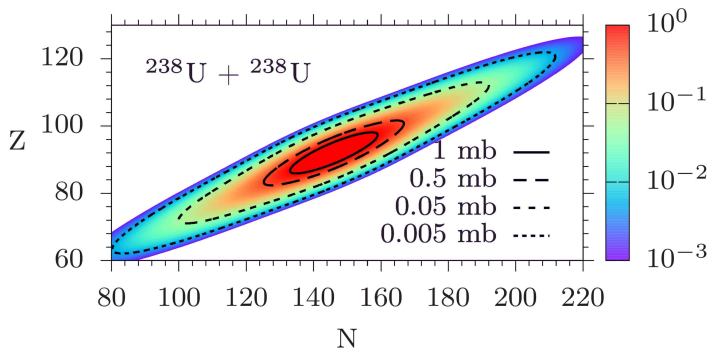}
	\caption{Double cross-sections $\sigma(N,Z)$ for production of primary fragments
		with neutron numbers N and proton numbers Z in (N,Z)-plane in the collisions of
		${}^{238} \text{U}+{}^{238} \text{U}$ at $E_\text{c.m.} =833$~MeV. Elliptic
		lines indicate the set of isotopes with equal production cross-sections.}
	\label{fig4}
\end{figure} 
Fig.~\ref{fig4} shows the double cross-sections $\sigma (N,Z)$ in the
$(N,Z)-$plane. We observe the cross-section distribution extends along the
bottom of the beta stability and exhibits large dispersion in this direction as
a result of the slight negative curvature of the potential energy along the
iso-scalar direction. We note that the nucleon diffusion along the beta
stability line is rather sensitive to the magnitude of the reduced iso-scalar
curvature parameter $\beta$. Decreasing the magnitude of this parameter, the
dispersion of the double cross-section along the beta stability direction is
reduced. The cross-sections $\sigma (A)$ as a function of the mass numbers of
the primary fragments are given by,
\begin{align} \label{eq21}
\sigma (A)=\frac{\pi \hbar ^{2} }{2\mu E_\text{c.m.} } \sum _{\ell_{\min } }^{\ell_{\max } }(2\ell+1)P_{\ell}  (A).
\end{align} 
Here $P_{\ell} (A)=\left(P_{\ell}^{t-t} (A)+P_{\ell}^{s-s} (A)\right)/2$ denotes
the mean value of the probability of producing a primary fragment with mass
numbers $A$ in the tip-tip and the side-side collisions with the initial angular
momentum $\ell$. These probabilities are determined by a simple Gaussian
functions,
\begin{align} \label{eq22}
P(A)=\frac{1}{\sigma _{AA} \sqrt{2\pi } } \exp \left[-\frac{1}{2} \left(\frac{A-\overline{A}}{\sigma _{AA} } \right)^{2} \right],
\end{align} 
where the mass dispersion is determined by $\sigma _{AA}^{2} =\sigma _{NN}^{2}
+\sigma _{ZZ}^{2} +2\sigma _{NZ}^{2}$ and the mean mass number as
$\overline{A}=238$. Fig.~\ref{fig5} shows the cross-sections as a function of
the mass numbers of the primary fragments in the tip-tip and the side-side
geometries and their mean values. We can calculate the cross-sections $\sigma
(Z)$ of production of the primary fragments as a function of the atomic number
using an expression similar to Eq.~\eqref{eq21} by employing the Gaussian
probability with the dispersion and the mean values as given by  $\sigma _{ZZ}
(Z)$ and $\overline{Z}=92$, respectively. Figure~\ref{fig6} shows the
cross-sections as a function of the atomic numbers of the primary fragments in
the tip-tip and the side-side geometries and their mean values.
\begin{figure}[!hpt]
\includegraphics*[width=8.6cm]{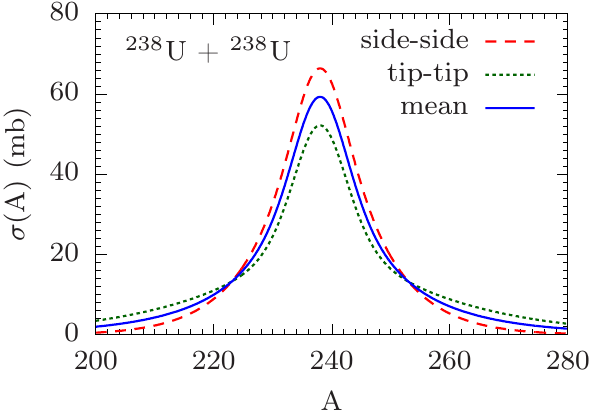}
\caption{Cross-sections $\sigma (A)$ for production of primary fragments as a
function of mass number in the collisions of ${}^{238} \text{U}+{}^{238}
\text{U}$ at $E_\text{c.m.} =833$~MeV in tip-tip, side-side geometries and  mean
values by dashed, dotted and solid blue lines, respectively.}
\label{fig5}
\end{figure} 
\begin{figure}[!hpt]
\includegraphics*[width=8.6cm]{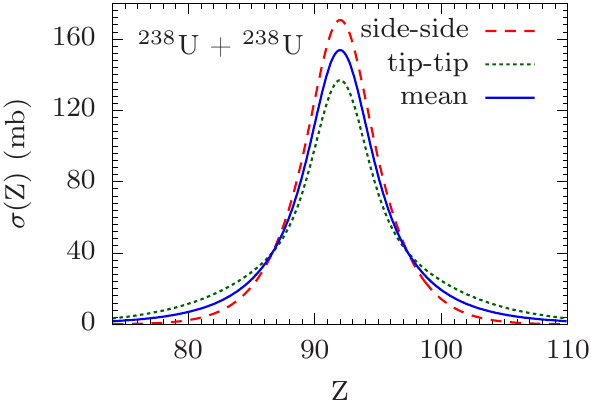}
\caption{Cross-sections $\sigma (Z)$ for production of primary fragments as a
function of mass number in the collisions of ${}^{238} \text{U}+{}^{238}
\text{U}$ at $E_\text{c.m.} =833$~MeV in tip-tip, side-side geometries and  mean
values by dashed, dotted and solid blue lines, respectively.}
\label{fig6}
\end{figure} 

\section{CONCLUSIONS}
\label{sec4}
We have carried out an investigation of mass and charge distributions of the primary
fragments produced in the collisions of the ${}^{238} \text{U}+{}^{238}
\text{U}$  system at $E_\text{c.m.} =833$~MeV.  We calculate the probability
distributions of the primary fragments by employing the quantal diffusion
description.  In the quantal diffusion approach, the concepts of the transport
theory are merged with the mean-field description of the TDHF with the help of
the SMF approach.  It is then possible to express the
diffusion coefficients of the relevant macroscopic variables in terms of the
occupied single-particle wave functions of the TDHF.  Since the Langevin
equations of the macroscopic variables are equivalent to the Fokker- Planck
description for the distribution of the macroscopic variables, under certain
conditions, it is possible give nearly analytical description for the
distribution functions of the macroscopic variables and the cross-sections. In
the calculations of the cross-sections of production of the primary fragment for
each initial angular momentum or equivalently for each impact parameter, we need
to determine the mean values of the neutron and proton numbers of the fragments
and the neutron, the proton and the mixed dispersions of the distribution
functions.  The mean values are determined by the TDHF descriptions. The
variances are calculated from the solutions of three coupled differential
equations in which diffusion coefficients of neutron and protons act as the
source terms. The behavior of the potential energy surface of the di-nuclear
complex makes an important effect on the neutron and proton diffusion
mechanism.  It is possible to determine the curvature parameters of the
potential energy in the collisions of asymmetric systems from the drift
information with the help of the Einstein's relation in the over-damped limit.
Since collisions of the symmetric systems, such as the collisions of ${}^{238}
\text{U}+{}^{238} \text{U}$, do not exhibit drift of the neutron and proton
degrees of freedom, we need to employ other methods to specify the curvature
parameters of the potential energy.  In this work, we employ the central
collision of a neighboring system ${}^{236} \text{Ra}+{}^{240} \text{Cm}$ at the
same bombarding energy.  The system initially drifts nearly along the iso-vector
direction and reach the charge equilibrium state rather rapidly.  From the
iso-vector drift information, we can estimate the reduced curvature parameter of
the potential energy as $\alpha =0.13$. After reaching the charge equilibration,
the system spends a long time in the vicinity of ${}^{238} \text{U}+{}^{238}
\text{U}$ state and eventually has a tendency drift along the iso-scalar path
away from the symmetric state. This behavior indicates the symmetric ${}^{238}
\text{U}+{}^{238} \text{U}$ is located at an unstable equilibrium position with
a small negative curvature toward the iso-scalar direction. However, from the
drift information it is not possible to estimate the iso-scalar reduced
curvature parameter $\beta$. Since the negative curvature may lead to broad
diffusion along the beta stability line, it is important to determine this
curvature parameter accurately. Therefore, we regard the reduced curvature in
the iso-scalar direction as a parameter and estimate its value with the help of
the isotopic cross-section data of gold nucleus from a previous investigation of
the ${}^{238} \text{U}+{}^{238} \text{U}$ collisions at about the same energy. 
In this work, we present calculations for production of the primary fragments
with the curvature parameters $\alpha =0.13$ and $\beta =-0.02$. The primary
fragments are excited and cool dawn by the de-excitation processes of particle
emission, mostly neutrons and by sequential fission of the heavy fragments.
Calculations of the secondary cross-sections exceed the scope of the present
work. We plan to investigate the de-excitation process of the primary fragments
in the collisions of ${}^{238} \text{U}+{}^{238} \text{U}$ and calculate the
secondary cross-sections in a subsequent study.
 
\begin{acknowledgments}
S.A. gratefully acknowledges the IPN-Orsay and the Middle East Technical
University for warm hospitality extended to him during his visits. S.A. also
gratefully acknowledges useful discussions with D. Lacroix, K. Sekizawa,
D. Ackermann, and very
much thankful to his wife F. Ayik for continuous support and encouragement. This
work is supported in part by US DOE Grants Nos. DE-SC0015513 and DE-SC0013847, 
and in part by TUBITAK Grant No. 117F109.
\end{acknowledgments}
 
\appendix
\section{CURVATURE PARAMETERS OF THE POTENTIAL ENERGY}
\label{appA}
\begin{figure}[!hpt]
	\includegraphics*[width=8.6cm]{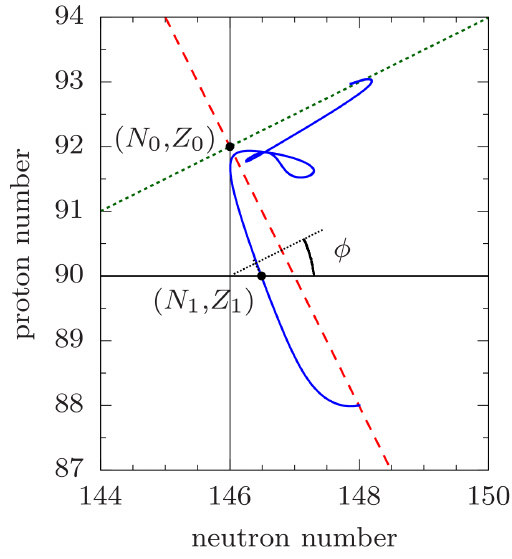}
	\caption{Drift path of the radium-like fragments in the central collisions of
		the ${}^{236} \text{Ra}+{}^{240} \text{Cm}$ system at $E_\text{c.m.} =833$~MeV
		(solid blue line) at tip-tip geometry. Dashed green line starting at $(Z_{1}
		=92,N_{1} =144)$ and the perpendicular dashed red line are the iso-scalar and
		iso-vector paths, respectively.}
	\label{figA1}
\end{figure}

The charge asymmetry of uranium ${}_{92}^{238} \text{U}_{146}$ is
$(146-92)/(146+92)=0.227$. The dashed green line in Fig.~\ref{figA1} represents
the nuclei with nearly equal charge asymmetry $(N-Z)/(N+Z)=0.22-0.23$. We refer
to this line as the iso-scalar line which extends nearly parallel to the lower part of
the beta stability valley in this region. We refer to the dashed red line as the
iso-vector line which is perpendicular to the iso-scalar path. We parameterize
the potential energy surface in the vicinity of the equilibrium $(N_{0}
=146,Z_{0} =92)$ in terms of two parabolic forms along the iso-scalar and
iso-vector paths as,
\begin{align} \label{a1}
U(N_{1} ,Z_{1} )=\frac{1}{2} aR_{S}^{2} (N_{1} ,Z_{1} )+\frac{1}{2} bR_{V}^{2} (N_{1} ,Z_{1}). 
\end{align}
The vertical distances $R_{S}$ and $R_{V}$ of a point ($N_{1} ,Z_{1} $)
representing a fragment from the iso-scalar and the iso-vector lines,
respectively are given by,
\begin{align} \label{a2}
R_{S} =&(Z_{0} -Z_{1} )\cos \phi +(N_{1} -N_{0} )\sin\phi,\nonumber\\ 
R_{V} =&(Z_{0} -Z_{1} )\sin \phi -(N_{1} -N_{0} )\cos \phi. 
\end{align}
According to the Einstein relation in the over-damped limit neutron and proton drift coefficients are related to the driving potential as,
\begin{eqnarray*}
v_{n} =-\frac{D_{NN} }{T} \frac{\partial U}{\partial N_{1} } =-\alpha D_{NN} R_{S} \sin \phi +\beta D_{NN} R_{V} \cos \phi\label{a3}\\
v_{z} =-\frac{D_{ZZ} }{T} \frac{\partial U}{\partial Z_{1} } =+\alpha D_{ZZ} R_{S} \cos \phi +\beta D_{ZZ} R_{V} \sin \phi\label{a4}\;.
\end{eqnarray*}
%\begin{align} \label{a3}
%v_{n} =-\frac{D_{NN} }{T} \frac{\partial U}{\partial N_{1} } =-\alpha D_{NN} R_{S} \sin \phi %+\beta D_{NN} R_{V} \cos \phi 
%\end{align}   
%\begin{align} \label{a4}
%v_{z} =-\frac{D_{ZZ} }{T} \frac{\partial U}{\partial Z_{1} } =+\alpha D_{ZZ} R_{S} \cos \phi %+\beta D_{ZZ} R_{V} \sin \phi\;.
%\end{align} 
Here, the temperature is absorbed in the reduced curvature parameters as $\alpha
=a/T$ and $\beta =b/T$. Because of the analytical form, we can readily calculate
the derivatives of the drift coefficients to obtain,
\begin{eqnarray}
\partial v_{n} (t)/\partial N_{1} =-D_{NN} \left(\beta \cos ^{2} \phi +\alpha \sin ^{2} \phi\right) \label{a5} \\
\partial v_{n} (t)/\partial Z_{1} =+D_{NN} \left(\alpha -\beta \right)\cos \phi \sin \phi \label{a6}\\
\partial v_{p} (t)/\partial Z_{1} =-D_{ZZ} \left(\beta \sin ^{2} \phi +\alpha \cos ^{2} \phi \right) \label{a7}\\
\partial v_{p} (t)/\partial N_{1} =+D_{ZZ} \left(\alpha -\beta \right)\cos \phi \sin \phi \label{a8}\;.
\end{eqnarray}
%\begin{align} \label{a5}
%\partial v_{n} (t)/\partial N_{1} =-D_{NN} \left(\beta \cos ^{2} \phi +\alpha \sin ^{2} %\phi\right) 
%\end{align}         
%\begin{align} \label{a6}
%\partial v_{n} (t)/\partial Z_{1} =+D_{NN} \left(\alpha -\beta \right)\cos \phi \sin \phi 
%\end{align}
%\begin{align} \label{a7}
%\partial v_{p} (t)/\partial Z_{1} =-D_{ZZ} \left(\beta \sin ^{2} \phi +\alpha \cos ^{2} \phi %\right) 
%\end{align}
%\begin{align} \label{a8}
%\partial v_{p} (t)/\partial N_{1} =+D_{ZZ} \left(\alpha -\beta \right)\cos \phi \sin \phi 
%\end{align}
The reduced curvature parameters are determined by the drift and the diffusion coefficients as,
\begin{align} \label{a9}
\alpha R_{S} (t)=\frac{v_{z} (t)}{D_{ZZ} (t)} \cos \phi -\frac{v_{n} (t)}{D_{NN} (t)} \sin \phi 
\end{align}
and
\begin{align} \label{a10}
\beta R_{V} (t)=\frac{v_{z} (t)}{D_{ZZ} (t)} \sin \phi +\frac{v_{n} (t)}{D_{NN} (t)} \cos \phi\;.
\end{align}

In collisions of symmetric systems, the drift coefficients vanish and the mean
values of the neutron and proton numbers of the fragments are equal to the
equilibrium values of the colliding nuclei $\overline{N}_{1} =N_{0}$,
$\overline{Z}_{1} =Z_{0}$. As a result, it is not possible to determine the
reduced curvature parameters from the Eq.~\eqref{a9} and Eq.~\eqref{a10}.  

\begin{figure}[!hpt]
\includegraphics*[width=8.6cm]{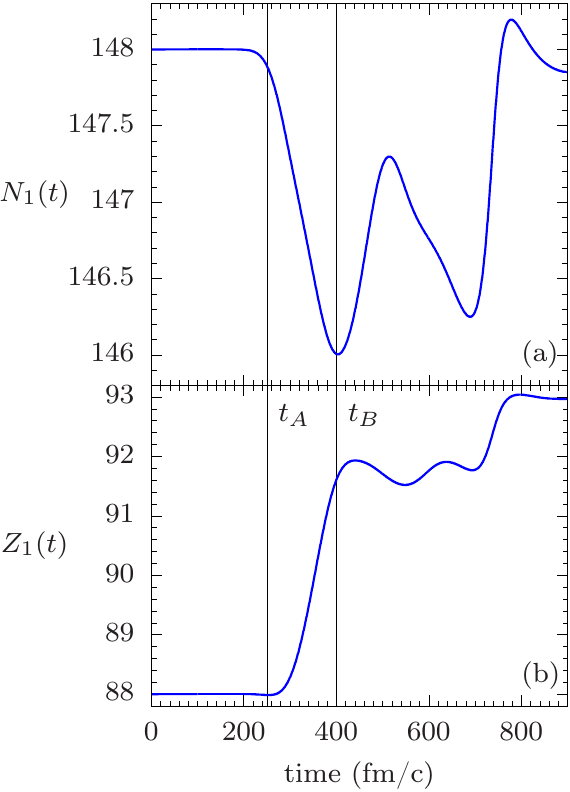}
\caption{Neutron $N_{1} (t)$ and proton $Z_{1} (t)$ numbers of radium-like
fragments as function of time in the central collisions of the ${}^{236}
\text{Ra}+{}^{240} \text{Cm}$ system at $E_\text{c.m.} =833$~MeV .(solid blue
line) at tip-tip geometry.}
\label{figA2}
\end{figure}

In order to estimate the reduced curvature parameters, we consider the central
collision of a neighboring system of ${}_{88}^{236} \text{Ra}_{148}
+{}_{96}^{240} \text{Cm}_{144}$ at the same bombarding energy $E_\text{c.m.}
=833$~MeV. We consider ${}^{236} \text{Ra}$ as the projectile. Figure~\ref{figA2}
shows the neutron mumber ${N}_{1}(t)$ and the proton number ${Z}_{1}(t)$
as a function of time. Blue line in
Fig.~\ref{figA1} shows the drift path of the projectile-like fragments in the $(N,Z)$-plane. We
observe that the system rapidly evolves toward the equilibrium charge asymmetry
of the ${}^{238} \text{U}+{}^{238} \text{U}$ system nearly along the iso-vector
direction from the initial state at point A toward the state at point B. This
segment of the drift path is suitable to determine the average value of the
reduced iso-vector curvature as,
\begin{align} \label{a11}
\alpha \int _{t_{A} }^{t_{B} }dt\left[R_{S} (t)\right] =\int _{t_{A} }^{t_{B} }dt\left[\frac{v_{z} (t)}{D_{ZZ} (t)} \cos \phi -\frac{v_{n} (t)}{D_{NN} (t)} \sin \phi \right] 
\end{align}
where $t_{A} =250$~fm/c and $t_{B} =400$~fm/c as indicated in Fig.~\ref{figA2}.  
We find the reduced iso-vector
curvature parameter as $\alpha =0.13$. In Fig.~\ref{figA1}, after the symmetric
state ${}^{238} \text{U}+{}^{238} \text{U}$, because of quantal effects due
to shell structure, the TDHF drift path follows a complex pattern for a long
time and subsequently appears to drift toward asymmetry along the iso-scalar direction. This
behavior indicates that the symmetric state is an unstable equilibrium point in
the iso-scalar direction, i.e. along the beta stability line, and the average
potential energy has an inverted parabolic shape with a negative curvature
parameter. Such potential shape may lead to relative large diffusion along the
beta stability direction. Unfortunately, the drift segment after the symmetric
state until the time which the fragment separates is not suitable to estimate the
average value of the reduced iso-scalar curvature parameter~$\beta$. 
\begin{figure}[!hpt]
\includegraphics*[width=8.6cm]{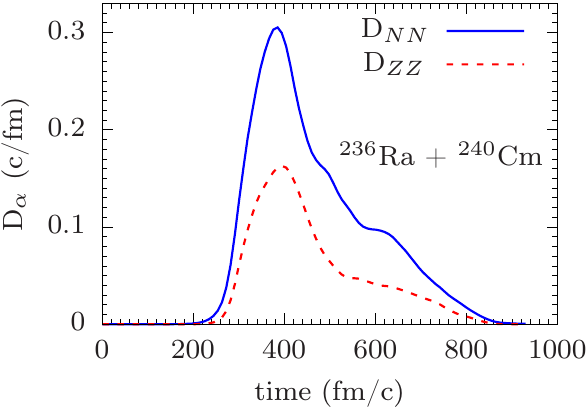}
\caption{Neutron and proton diffusion coefficients as function of time in the
central collisions of the ${}^{236} \text{Ra}+{}^{240} \text{Cm}$ system at
$E_\text{c.m.} =833$~MeV at tip-tip geometry.}
\label{figA3}
\end{figure} 

\section{CURVATURE PARAMETERS ALONG THE BETA STABILITY}
\label{appB}
\begin{figure}[!hpt]
\includegraphics*[width=8.6cm]{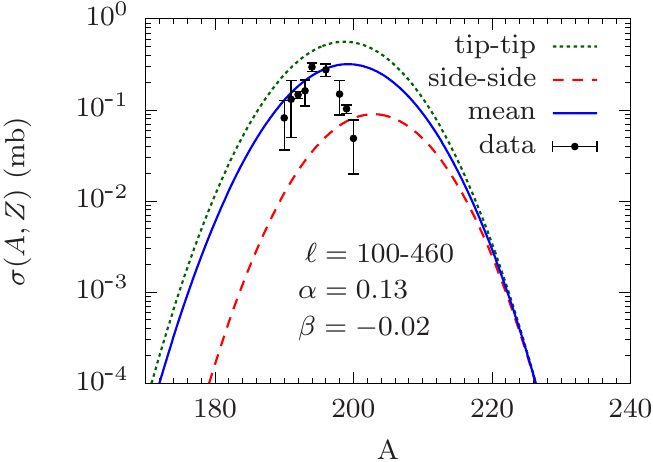}
\caption{Cross-section of gold $Z=79$ isotopes averaged over tip-tip and
side-side geometries as a function of the mass$A$numbers in the collisions of
${}^{238} \text{U}+{}^{238} \text{U}$ system at $E_\text{c.m.} =833$~MeV
calculated with curvature parameters $\alpha =0.13$ and $\beta =-0.02$. Solid
dots indicate data taken from~\cite{kratz2013}.}
\label{figB1}
\end{figure}

We consider the reduced iso-scalar curvature $\beta$ as a parameter. In order
obtain a reasonable value for $\beta$, we employ the cross-section data for
production of gold isotopes from a previous investigation of the ${}^{238}
\text{U}+{}^{238} \text{U}$ system at about the same energy.  We calculate the
distribution of the cross-sections $\sigma (N=A-Z,Z)$ primary gold isotopes with
the atomic number $Z=79$ and mass numbers $A$ using Eq.~\eqref{eq20} for the
double $\sigma (N,Z)$ cross-sections. In order to cover the angular range of the
experimental set up in Ref.~\cite{kratz2013}, the range of the angular momentum
summation in Eq.~\eqref{eq20} is taken as  $\ell_{\min } =100$ and $\ell_{\max }
=460$. Fig.~\ref{figB1} shows the cross-sections for production of the primary
gold isotopes which are calculated with the reduced iso-vector curvature $\alpha
=0.13$ and the reduced iso-scalar curvature $\beta =-0.02$. The primary gold
isotopes are excited and cool down mainly by neutron emissions.  In determining
the average number of the emitted neutrons, we need to estimate the average
excitation energy of these isotopes.  The TDHF calculations presented in 
Table~\ref{tab1} and Table~\ref{tab2}, do not give accurate information for the total
kinetic energy loss (TKEL) in these channels. However for a rough estimate we
can take the results for the initial angular momentum $L=300\hbar$, which is
about the waited mean value of the angular momentum range. For this angular
momentum, the TKEL in the tip-tip and the side-side geometries are $217$~MeV and
$135$~MeV, respectively. For the gold channel U+U $\rightarrow$
Au(195,79)+Db(281,105) the $Q_{gg}-$value is 24.1~MeV. Sharing the TKEL and the
$Q_{gg}-$value in proportion to the masses, we find the average excitation
energy of the gold isotopes to be $86.8$~MeV and $54.0$~MeV, in the tip-tip and
the side-side geometries, respectively.  Assuming one neutron emitted per
$10.0$~MeV, on the average about $9$, $5$ and $7$ neutrons are emitted in the
tip-tip, in the side-side and in the mean geometry, respectively. In
Fig.~\ref{figB1}, if we shift the mean gold isotope distribution by~7 units to
the left, the peak value of the cross-sections matches the peak value of the
gold data. This indicates $\beta =-0.02$ is a reasonable estimate for the reduced
curvature parameter in the iso-scalar direction. We note that the calculations
overestimate the isotopic width, which most probably is due to the parabolic
approximation of the potential energy.

\bibliography{VU_bibtex_master}
\end{document}